\documentclass[12pt]{iopart}

\usepackage{graphicx}
\begin{document}

\title[Inverse Relationship Between Molecular Diversity And Resource Abundances]{Inverse Relationship Between Molecular Diversity And Resource Abundances}

\author{Atsushi Kamimura \& Kunihiko Kaneko}

\address{Department of Basic Science, The University of Tokyo, 3-8-1, Komaba, Meguro-ku, Tokyo 153-8902, Japan}
\ead{kamimura@complex.c.u-tokyo.ac.jp}
\vspace{10pt}

\begin{abstract}
Cell reproduction involves replication of diverse molecule species, in contrast to simple replication system with fewer components. 
Here, we address why such diversity is sustained despite the efficiency of simple replication systems, using a cell model 
with catalytic reaction dynamics that grew by uptake of environmental resources. 
Limited resources led to increased diversity of components within the system, and the number of coexisting species increased with a negative power of the resource abundances. 
The diversity was explained from the optimum growth speed of the cell, determined by a tradeoff between the utility of diverse resources and the concentration onto fewer components to increase the reaction rate.
\end{abstract}

%
%
%
%
%

\section{Introduction}

Diverse molecule species coexist in a cell; these components are encapsulated in cellular compartments and synthesized with the aid of catalysts in order to achieve cell reproduction by taking up 
essential nutrients and resources from the environment. 
Indeed, huge diversity in components is achieved. However, when considering the theoretical minimum requirements of a cell, a simple cell consisting of fewer components has been 
shown to achieve a more rapid growth speed\cite{Protocell}. Importantly, the collision rate of a molecule with its catalyst will be increased if the composition of the cell 
includes fewer species that catalyze the replication of one another. 
For example, in cell models consisting of a catalytic reaction network\cite{EigenSchuster, Eigen, Dyson, Kauffman, Jain, Lancet, Furusawa, Kaneko, KamimuraKanekoPRL2010}, 
an autocatalytic set is formed in which a collection of molecular species that synthesizes themselves through catalysis by other molecules within the set, 
the presence of fewer components has been shown to achieve more rapid growth.
Several $in$ $vitro$ and $in$ $silico$ models have supported these findings\cite{Spiegelman, Fontana, Ray}. In general, cells with simple reaction pathways consisting of only 
a few molecular species would achieve a higher reproduction rate than a complex system containing diverse components, 
and the strategy to increase compositional diversity with multiple pathways would be evolutionarily selected out.
However, in the present cells, there is a huge diversity of molecules. Thus, a theoretical explanation of this diversity is needed in order to improve our understanding of biological systems.

In the experiments and theoretical models for the cells, it is implicitly assumed that resources for the synthesis of molecules are sufficiently abundant.  
As long as the resources are sufficiently available, 
simplification of the molecular components would be more favorable for reproduction than diversification. 
In contrast, if resources are limited, focusing on only a few components using specific resources may not be evolutionarily stable. 
Thus, because of resource limitation, cells with diverse components and multiple pathways that manage various resources to achieve growth would be more evolutionarily favorable. 
In fact, the presence of multiple pathways due to diversity of components stabilizes an evolving network\cite{Jain2} by ensuring that there are alternative routes to achieve reproduction.
This is analogous to the selection theory of ecological population in which there is a tradeoff between increasing the intrinsic growth rate and the carrying capacity\cite{rK}. 

However, there is an important difference between molecule replication in a cell and individual replication in an ecosystem.
A cell, an ensemble of molecules, reproduces itself, competes for growth, and is a unit for
the selection, whereas ecosystem itself is not an object for selection.
Here we are interested in a hierarchic growth system, in which cell growth and molecule replication have to be balanced.  
Indeed, we will show that such a balance leads to the diversity transition in molecular components with the decrease in resource abundances
as well as a general scaling law between the diversity and resource abundances.

In the present paper, we study the possible relationships between compositional diversity and resource abundances, by investigating a simplified cell model consisting of 
a catalytic reaction network in which each molecule replicates by consuming respective resources. 
We first demonstrate that there is a certain threshold for resource abundances below which the compositional diversity, i.e., the number of coexisting chemical species is increased.   
This increase follows the negative power law of resource abundances. Importantly, we could explain this scaling relationship 
by the tradeoff between the increase in components for the use of diverse resources and the decrease in components to increase the reaction rate. 

\section{Model}

For a cell to reproduce itself, all catalytic components have to be synthesized with the aid of catalysts.
Thus a catalytic component $i$ is replicated with the aid of $j$, while the component $j$ is replicated with the aid of some other component. 
Following this general requirement, Eigen introduced the hypercycle model, which consists of catalytic reaction $X_j + X_i \rightarrow 2 X_j + X_i$\cite{EigenSchuster}.  
Considering networks of such reactions, Kauffman introduced the concept of autocatalytic set, while a protocell model with such reaction networks has been investigated\cite{Kauffman}.   
In this general scheme for a cell model with replicating molecules, however, resource molecules are not explicitly included, or in other words, resources are fully supplied.  

\begin{figure}[t]
\begin{center}
\includegraphics[width=10cm]{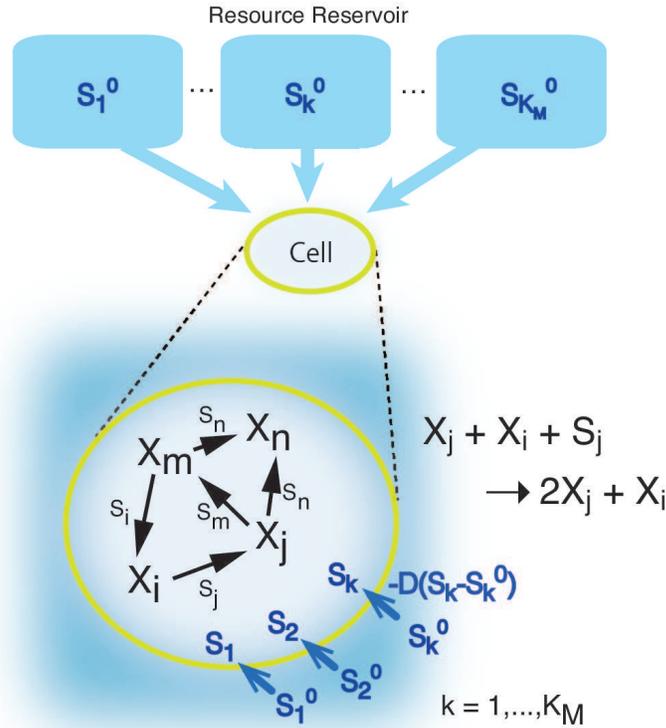}
\vskip 0.25cm
\caption{Schematic representation of our model.
The system is composed of $M_{\rm tot}$ cells, each of which contains molecule species
$X_j$ $(j=1,...,K_M)$ forming a catalytic reaction network 
to replicate each $X_j$. Each cell takes up resources $S_k$$(k=1,...,K_M)$ to consume each for replicating $X_k$ 
from the resource reservoir in the environment via diffusion $-D(S_k - S_k^0)$, 
where $S_k^0$ is a randomly-fixed constant $S_k^0 \in [0,10]$ and $D$ is the diffusion constant.  }
\label{fig:1}
\end{center}
\end{figure}

Here we modified the standard hypercycle model to include the resources.  
We adopted a cell model in which each molecule ($X_j; $$j=1,..,K_M$) was replicated, 
by consuming a corresponding resource ($S_{j};$$j=1,...,K_M)$(see Fig. \ref{fig:1}).
$M_{\rm tot}$ cells were defined; each contains $K_M$ species of replicating molecules where some of species possibly have a null population. 
Molecules of each species $X_j$ were replicated with the aid of some other catalytic molecule $X_i$, determined by a random catalytic reaction network, by consuming a predetermined resource $S_{j}$, 
one of the supplied resource chemicals $S_k$$(k=1,...,K_M)$, as follows:
\begin{equation}
X_j + X_i + S_{j} \rightarrow 2X_{j} + X_i.
\label{reaction}
\end{equation}
For this reaction to replicate $X_j$, one resource molecule is needed, and 
the replication reaction does not occur if $S_{j}$ is less than one. 
The reaction coefficient is given by the catalytic activity $c_i$ randomly determined as $c_i \in [0,1]$ of the molecule species $X_i$.
With each replication, error occurs with probability $\mu$.
This error corresponds to changes in some monomers in the polymer sequence and catalytic properties of the molecule. 
Here, for simplicity, for each replication of $X_j$, the molecule is replaced by a different molecule $X_l$ ($l \neq j$) with equal probability $\mu/(K_M-1)$ 
where $K_M$ is the number of molecule species.

Each cell takes up resources ($S_j$) by diffusion from its respective resource reservoir.
From external reservoirs of concentrations $S_j^0$, the resources ($S_j$) are supplied into each cell by diffusion $-D(S_j - S_j^0)$.
$D$ controls the degree of the uptake rate because the resource supply is reduced by decreasing $D$.
We carried out stochastic simulations of the model, as detailed in Appendix.

A random catalytic reaction network was constructed as follows.
For each molecule species, the density for the path of the catalytic reaction was 
given by $\rho$(which was fixed at 0.2), such that each species had $\rho K_M$ reactions on 
average. Once chosen, an identical reaction network was adopted during each simulation for all cells. 
Even though the underlying network did not change, each cell used only a part of the reaction pathways, 
depending on its composition(e.g., both $X_i$ and $X_j$ must be present in the cell for reaction (\ref{reaction}) to occur). 
Autocatalytic reactions in which $X_i$ catalyzed the replication of itself were excluded from the catalytic network, 
and direct mutual connections were also excluded such that $X_j$ did not function as a catalyst for $X_i$ if $X_i$ was the catalyst for $X_j$.

When the total number of molecules in each cell exceeded a given threshold $N$, 
the cell divided into two cells and the molecules were randomly partitioned. 
One randomly chosen cell was removed from the system in order to fix the total number of cells at $M_{\rm tot}$.

\section{Results}

\subsection{Diversification of composition}
\begin{figure}[t]
\begin{center}
\includegraphics[width=9cm]{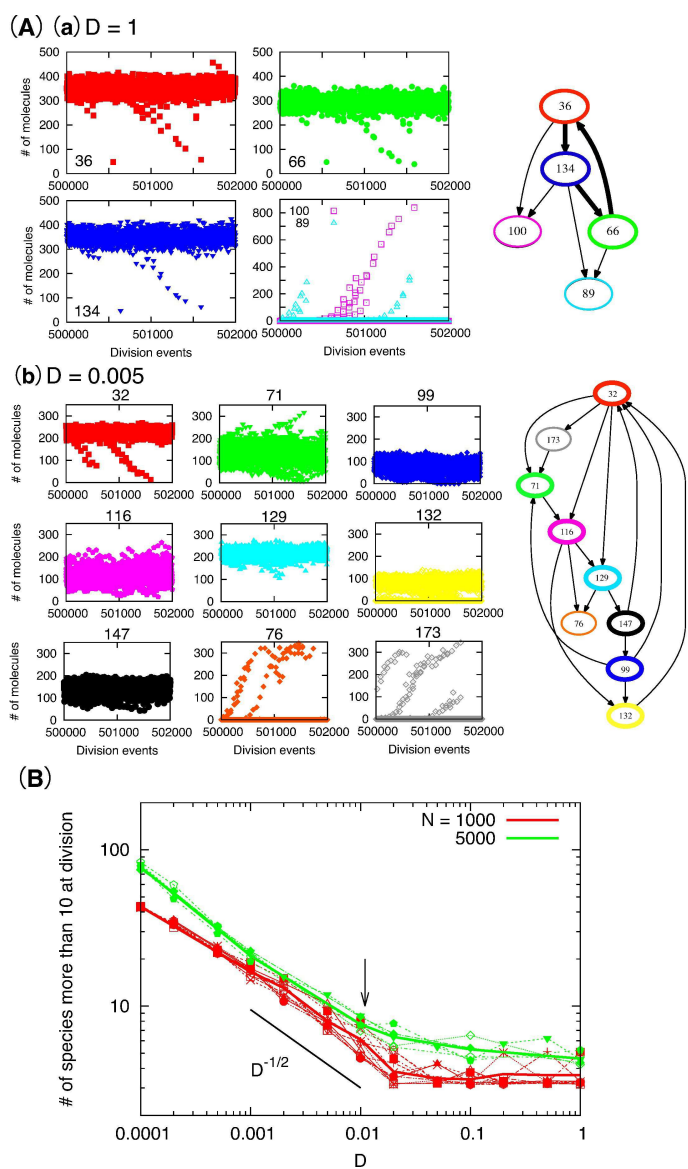}
\vskip 0.25cm
\caption{(A) The major composition of cells for (a)D = 1 and (b)D = 0.005. The number of molecules for the species at division events is shown for 2000 successive division events in the system. 
(a) For D = 1, the three molecule species ($36$, $66$ and $134$) dominated the composition (each approximately with 1/3 of $N=1000$), and
almost all dividing cells had the same composition. However, some cells were dominated by parasitic molecules (species $89$ and $100$) and could not survive.
The right panel shows the catalytic network formed by three species (the number indicates the species $i$, and the arrow from species $i$ to $j$ indicates that $X_i$ was a catalyst for replicating $X_j$. 
(b) For D = 0.005, more species were present in the cells, forming the larger network shown in the right panel. 
Almost all cells had similar compositions, while some were dominated by parasitic species (species$76$ and $173$).
The parameters were as follow: $K_M = 200$, $M_{\rm tot} = 100$, $N = 1000$, and $\mu = 0.001$.
(B) The number of major species (more than 10 copies at division) as a function of $D$ for $N = 1000$ and $5000$.
Thin lines with points and thick curves show results of different network samples and their averages, respectively. 
The estimated balance point is shown by the arrow. The slope $D^{-1/2}$ is also shown as a visual guide.
}
\label{fig:2}
\end{center}
\end{figure}

We investigated how diversity in composition changed with the uptake rate of the resources $D$. 
When the resources were supplied at a sufficiently rapid rate (e.g., for $D = 1$), 
three components typically dominated most of the composition(each representing approximately 1/3 of the molecule population; see Fig.\ref{fig:2}(A)(a)). 
Thus, the minimum hypercycle was formed by three components\cite{Eigen,EigenSchuster} as shown in the right panel of Fig. \ref{fig:2}(A)(a). 
The hypercycle established a recursively growing state, where the composition was robust against stochasticity in reactions and perturbations by the division processes. 
Most of the other molecular species were absent, while some species appeared due to replication error.
Some {\sl parasitic} species that were catalyzed by a member of the hypercycle but did not catalyze other members were found to increase in number on occasion\cite{MaynardSmith, Szathmary, McCaskill,Hogeweg}. 
However, cells dominated by the parasitic molecules could not continue growth(see Fig. 1 in \cite{Supplementary}).
All dividing cells adopted this three-component hypercycle, and there was no compositional diversity; cells use the minimum reaction pathway to grow.

As $D$ decreases below a certain threshold $D_c\sim 0.01$, the number of molecular species increased, and multiple reaction pathways were utilized(Fig.\ref{fig:2}(A)(b)).
Similar to the three-component hypercycle, the molecular species in this case also formed a mutually catalytic hypercycle, shown in Fig. \ref{fig:2}(A)(b), such that 
every species in the network could be replicated with the aid of other species in the network.
All dividing cells adopted approximately the same compositions. Moreover, cells dominated by parasites appeared on occasion, but could not survive (see Fig. 2 in \cite{Supplementary}).

To quantitatively investigate this increase in the number of species, we show in Fig. \ref{fig:2}(B) the number of major species(more than 10 copies at division) as a function of $D$ by using different underlying networks. 
Irrespective of the network samples, the number transits to increase at $D_c = 0.01 - 0.02$, and increases $\approx D^{-1/2}$ as $D$ decreases below this threshold.
In contrast, for $D>D_c$, cells were mainly composed of just three primary molecular species.

This transition was estimated by determining the point where the consumption rate of resources by the intracellular reactions reached the maximal inflow rate.
Beyond this critical point, three species typically formed the hypercycle, with each species representing approximately 1/3 of the molecule population.
Within this system, the probability that a species $X_j$ encountered with its catalyst $X_i$ was approximately $1/9$, and the reaction can occur with a rate $c_i/9$. 
On the other hand, the maximum inflow rate of resources was $DS_j^0$.
Thus, the balance point was estimated as $D^* = c_i/9S_j^0$. 
In our simulation setup, the $c_i$ was set at $[0,1]$, but as cells with higher growth exhibited improved survival, 
the $c_i$ for molecules present in cells was shifted to a higher range$(\approx 0.8$; see Fig. 3 in \cite{Supplementary}). 
Likewise, $S_j^0 \in [0,10]$ was shifted to higher values$(\approx 7-8$; Fig. 4 in \cite{Supplementary}).
Hence, the critical point could be approximated as $D^* = 0.011$, consistent with the transition in our simulation results.

\subsection{Coexistence in simple mutually-catalytic reactions}
The transition to diversity with multiple pathways was then analyzed in terms of dynamical systems. 
As an illustration, we considered simpler reactions where two sets of mutually catalytic molecules $X$, $Y$, and $Z$, $Y$ were initially within the cells. 
The molecular species mutually catalyzed the replication of each other to form a minimal hypercycle as follows:
\begin{equation}
X + Y + S_X \rightarrow 2X + Y, \hspace{5mm} Y + X + S_Y \rightarrow 2Y + X,
\label{XY}
\end{equation}
and 
\begin{equation}
Z + Y + S_Z \rightarrow  2Z + Y, \hspace{5mm} Y + Z + S_Z \rightarrow 2Y + Z.
\label{ZY}
\end{equation}
We denoted the intrinsic catalytic activities of $X$, $Y$, and $Z$ as $c_X$, $c_Y$, and $c_Z$, respectively. 
Each reaction to replicate $X$, $Y$, and $Z$ consumed the resources $S_X$, $S_Y$, and $S_Z$, respectively.

\begin{figure}[t]
\begin{center}
\includegraphics[width=10cm]{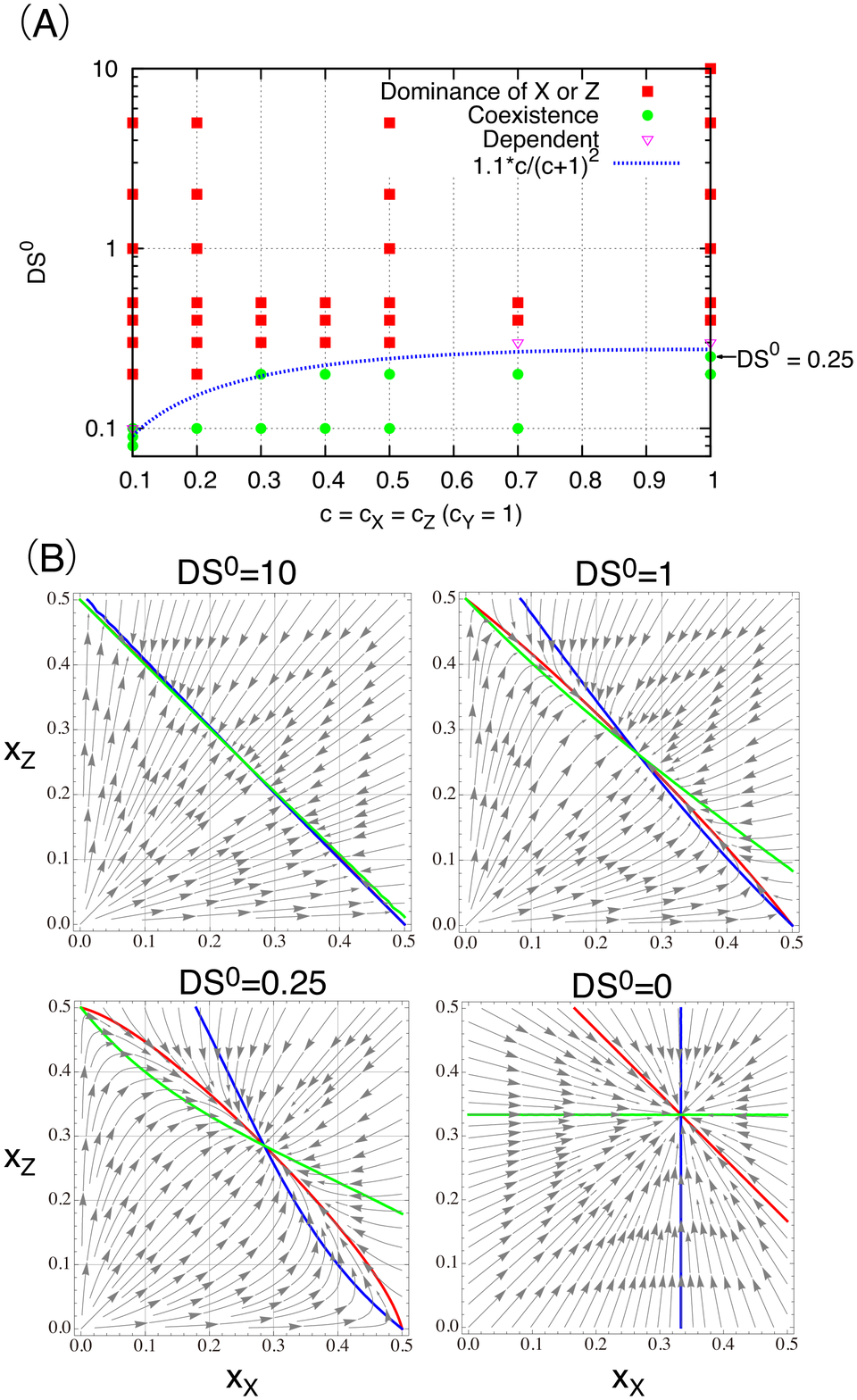}
\vskip 0.25cm
\caption{(A) Dominance and coexistence in the simple $X-Y$ and $Z-Y$ case as functions of $D$ and $c = c_X=c_Z$. $c_Y$ was fixed as 1. 
Points with dependent indicate that the outcome depends on the samples. Stochastic simulations were performed with same setups of the original model, except that only three species ($X, Y$, and $Z$) 
were considered and underwent the reactions (\ref{XY}) and (\ref{ZY}). (B) Nullclines in $x_X-x_Z$ plane for concentrations $x_Y$(Red), $x_X$(Blue) and $x_Z$(Green). Flows are also indicated by arrows. Here, $c_X = c_Y = c_Z = 1, S_X^0 = S_Y^0 = S_Z^0 = S^0 = 1$, $V = 1$ and $x_X + x_Y + x_Z = 1$. }
\label{fig:3}
\end{center}
\end{figure}

The results of stochastic simulations (Fig. \ref{fig:3}(A)) showed that transitions from selection of either $X$ or $Z$ to coexistence of $X$ and $Z$, together with $Y$, occurred with decreasing $D$. 
For example, by setting $c = c_X = c_Z$, we showed that the transition point was approximately equal to the balance point where $DS^0 \approx c_{Y} x_{X(Z)} x_Y$,  where $x_i$ for $i=X,Y,Z$ was 
the concentration of the corresponding component. 
When the resources were sufficiently abundant, either $X$ or $Z$ was selected in a steady state, given by $x_{X(Z)} = c_Y/(c+c_Y)$ and $x_Y = c/(c+c_Y)$. 
Then, the consumption rate of the resource $S_{X(Z)}$ was given as $c_Y x_{X(Z)} x_Y = c c_Y^2/(c + c_Y)^2$. 
Thus, the transition point was estimated as $D^* \approx c/(S^0(1+c/C_Y)^2)$, which agreed well with the simulation(see the dotted curve in Fig. \ref{fig:3}(A)).

This transition was also clarified by changes in the flow and nullclines in dynamical systems of rate equations for $c_X,c_Y,c_Z$ and resources, as shown in Fig. \ref{fig:3}(B) and described in 
$\S$1 of \cite{Supplementary}. 
To analyze this system using rate equations, we assumed that volume of the system was constant and that the total number of molecules was fixed.
Then, the increase in molecular species $Y$ could be written as 
\begin{equation}
\frac{dx_Y}{dt} = (c_X x_X x_Y + c_Z x_Y x_Z) S_Y - x_Y \phi, 
\end{equation}
and the increases for $X$ and $Z$ could be written as, 
\begin{equation}
\frac{dx_X}{dt} = c_Y x_X x_Y S_X - x_X \phi, \frac{dx_Z}{dt} = c_Y x_Z x_Y S_Z - x_Z \phi,
\end{equation}
where $x_I$ is the concentration of species $I$($X, Y, Z$), and $\phi$ is introduced to fix the total number of molecules, such that $\phi=c_Y x_Y \left( x_X S_X + x_Z S_Z \right) + 
\left( c_X x_X + c_Z x_Z \right) x_Y S_Y$.
By substituting the steady-state value of each resource into the rate equations, the nullclines and flow in the $x_X$-$x_Z$ plane were obtained as shown in Fig. \ref{fig:3}(B).

For sufficiently large values of $D$, the nullclines for $X$, $Y$, and $Z$ merged to a single line, and 
the solution was neutral on the line. In this case, even if both $X$ and $Z$ were initially present, the system converged on either $(x_X, x_Z) = (1/2, 0)$ or $(0,1/2)$ by fluctuations 
due to stochasticity in the reaction: thus either $X$ or $Z$ remained.
As $D$ was decreased, the nullclines split and crossed at a single point. This point corresponded to coexistence of both $X$ and $Z$, and 
this bifurcation occurred at $DS^0 = c_Y x_X x_Y$$(\sim 0.25$ in Fig. \ref{fig:3}(B)), consistent with our estimates thus far.

\subsection{Scaling behavior and optimum number of molecule species}
Below the transition to coexistence of multiple paths, further resource limitation, 
as shown in Fig. \ref{fig:2}(B), increased the compositional diversity with a negative power of $D$.
To explain this scaling relationship, 
we noted the existence of the following tradeoff: increasing the diversity in species enabled cells to utilize more resource species for their own growth, 
but decreased the reaction rate resulting from the collision of molecules with their catalysts. 
This tradeoff yielded the optimum number of remaining molecule species.

We estimate the optimal value for the remaining species as $K^{*}_M(\leq K_M)$, which achieved maximal growth under conditions of resource limitation. 
Here, we assumed that $K_M$ was sufficiently large to assure that $K^*_M$ could be increased to reach the optimum value.
Considering that a fixed set of $K^*_M$ molecule species mutually catalyzed the replication of each other, the temporal evolution of $N_i$, the number of a species $X_i$, was written as
$\frac{dN_i}{dt} \sim x_i x_j S_i$. 
Assuming that the concentrations of $K^*_M$ molecules were approximately the same, 
the concentration $x_{i(j)}$ was proportional to $\sim 1/K^*_M$. 
Therefore the increase in the number of molecules depended on the number of remaining species $K^*_M$s as follows:
\begin{equation}
\frac{dN_i}{dt} \sim \frac{1}{K^{*2}_M} S_i.
\label{dN}
\end{equation}

In the steady-state, the resource $S_i$ has a value $\bar{S_i}$ defined as 
\[ \bar{S_i} = \frac{DS_i^0}{1/K^{*2}_M + DS_i^0}. \]
If we assume that $S_i^0 = S^0$ for all $i$, 
the growth rate $G$ of the cell was defined as 
\[ G = \sum_i \frac{dN_i}{dt} \sim \frac{K^*_M D S^0}{1+ DS^0 K^{*2}_M}. \]
Given $D$ and $S^0$, the optimum value $K^{\rm opt}_M$ was obtained from  $dG/dK^*_M = 0$ as $K^{\rm opt}_M=(DS^0)^{-1/2}$.
Hence, as long as $K_M$ was sufficiently large to allow the above optimal value to be obtained, the scaling relationship $(DS^0)^{-1/2}$ was obtained, consistent with Fig. \ref{fig:2}(B).
\footnote{Note that the nonlinearity of the catalytic reaction, $x_i x_j$, was essential for determining the optimum.
In the linear reaction, i.e., $dN_i/dt \sim \frac{1}{K^*_M} S_i$, the $K^{\rm opt}_M$ may diverge such that the diversity would be increased as much as possible, i.e., $K_M$. }

For the estimate in eq. (\ref{dN}), we have excluded the possibility that each molecular species may have more catalyst species with the increase in the number of remaining species.
Including this possibility, eq. (\ref{dN}) was replaced by $\frac{dN_i}{dt} = \sum_j x_i x_j S_i$, where the summation runs over all the catalysts $X_j$ for $X_i$ in the present $K^*_M$ species.
Actually, the average number of catalysts for major molecules gradually increased as the resource was limited further(see Fig. 6 in \cite{Supplementary}). 
Further corrections to the scaling relationship could be needed for much smaller values of resources(see $\S$2 of \cite{Supplementary})
\footnote{
The estimation also assumed that the abundance of major molecule species were approximately equal, with a narrow distribution around a common value. 
This assumption is reasonable in our model because 
all the catalytic activities($c_i$) and resource abundance($S_i^0$) were of the same order; thus the number of each molecular species was approximately of the same order.
However, if the abundance is more broadly distributed, e.g., by the power-law distribution using a different setup\cite{Furusawa}, the scaling exponent could be modified.}.

As an example that did not require the above correction for the catalyst number, we considered a one-dimensional ring of the mutually catalytic reaction, 
\begin{equation*}
X_i + X_{i+1} + S_i \rightarrow 2X_i + X_{i+1}, \hspace{3mm} X_{i+1} + X_i + S_{i+1} \rightarrow  2X_{i+1} + X_i,
\end{equation*}
$(i = 1,...,K_M)$ with periodic boundary, i.e., $X_{K_M+1}$ denotes $X_1$.
In this case, the number of catalysts is two($X_{i-1}$ and $X_{i+1}$) for each $X_i$, irrespective of the number of present species in the cell.
The number of species increased clearly with $D^{-1/2}$ below the balance point $D^* = c_i/4S_j^0$(see $\S$3 and Fig. 5 in \cite{Supplementary}). 

\section{Summary and discussions}
In summary, we showed that diversification of compositions occurred as a result of limitation of various resources, when
the maximum inflow and consumption of resources were balanced. Using simple reactions, the transition was also clarified by
changes of nullclines. In addition, a gradual increase $\sim D^{-1/2}$ in the number of molecular species 
was explained by estimating the optimum number of species to give the maximum growth speed of cells.

Although we used a cell model consisting of hypercycle networks, our `diversity transition' is expected to be general for a cell system in which 
each component in a cell is replicated for its growth as a result of catalytic reactions, by consuming external resources. 
Then with the decrease in resource abundances, the diversity in intracellular components is increased.  Hence, the origin of diverse components in a cell is explained.

Our study provided a first step to explaining how replicating entities of the catalytic reaction network model respond to limitations and diversify their composition. 
The importance of diversity even at the primitive stage of life has been emphasized by Dyson\cite{Dyson}, while our study suggests the role of multiple resources in the intermediate stages from molecules to ecological population.
By corresponding replication of each molecule species with biological species in an ecosystem, the present diversification might have some similarity with
the studies in species diversification\cite{Gause, Tilman, Tilman2}. 
In spite of similarity, there is one important difference.  
In our study, cells, as an ensemble of molecules, reproduce and those with higher growth speed will be selected, whereas ecosystem itself is not a unit for reproduction and selection. 
Thus there is no direct pressure for a simple system with a higher reproduction rate.
In our study, both molecules and cells are units for reproduction and selection.
We expect that diversity transition with the decrease of resources is a general nature of such hierarchic reproduction systems.

In this study, we examined the compositional diversity of cells based on limitation of resources. 
However, competitions among cells may give rise to additional diversification; cellular phenotypes.
Indeed, different types of cells utilizing different sets of molecular species have developed to allow cell growth to occur.
In the case, the growth speed of cells and variations in cell compositions, allowing the cells to utilize different resources, will be also relevant for survival\cite{KamimuraKaneko2015}.


\section*{Appendix}
\subsection*{Simulation methods}

Simulations were carried out as follows. We introduced discrete simulation steps, as detained below.
For each simulation step, we repeated the following procedures. For each cell $q$ $(q=1,...,M_{\rm tot})$, 
we chose two molecules from the cell. If the pair of molecules, 
$X_i$ and $X_j$ were a catalyst and a replicator($X_j$ catalyzes the replication of $X_i$), the reaction occurred with the given probability($c_j$), 
if $S^q_i \geq 1$. $S^q_i$ denotes the resource to replicate $X_i$ assigned to each cell $q$. 
When the reaction occurred, the new molecule of $X_i$ was added into the cell and one molecule of the corresponding resource 
was subtracted to make $S^q_i \rightarrow S^q_i-1$. 
Here, with a probability $\mu$, a new molecule of $X_l (l \neq i)$, instead of $X_i$, was added into the cell, resulting in a structural change. 
If the total number of molecules in a cell exceeded the threshold $N$, the molecules were distributed into two daughter cells, while one cell, randomly chosen, was removed from the system. 
We also updated each $S^q_a$ to $S^q_a-D(S^q_a-S_a^0) (a = 1,...,K_M)$.

\section*{Acknowledgements}
  This work was supported by the Japan Society for the Promotion of Science. This work was also supported in part by the Platform for Dynamic Approaches to Living System from the Ministry of Education, Culture, Sports, Science, and Technology of Japan, and the Dynamical Micro-scale Reaction Environment Project of the Japan Science and Technology Agency.

\end{document}


\title[ ]{Supplementary material for \\Inverse Relationship Between Molecular Diversity And Resource Abundances}

\author{Atsushi Kamimura \& Kunihiko Kaneko}

\address{Department of Basic Science, The University of Tokyo, 3-8-1, Komaba, Meguro-ku, Tokyo 153-8902, Japan}
\ead{kamimura@complex.c.u-tokyo.ac.jp}
\vspace{10pt}


%
%
%
\maketitle

\section{A simple case - two sets of mutually catalyzing molecules -}
Let us consider three molecule species $X$, $Y$, and $Z$.
They replicate through the following reactions, 
\[ X + Y + S_X \rightarrow 2X+Y, \hspace{5mm} Y + X+ S_Y \rightarrow 2Y+ X \]
and 
\[ Y + Z + S_Y \rightarrow 2Y+Z, \hspace{5mm} Z + Y + S_Z \rightarrow 2Z+ Y. \]
We denote the intrinsic catalytic activities of $X$, $Y$, and $Z$, as $c_X$,  $c_Y$ and $c_Z$, respectively.
Each reaction to synthesize $X$, $Y$, and $Z$ utilized the resource $S_X$, $S_Y$, and $S_Z$, respectively.

The intrinsic reaction rates for replicating the molecule species $Y$ are given by
\[ F_Y = V (f_Y^X+f_Y^Z) s_Y, \]
respectively. 
Here, $V$ is the volume of the system, and 
$f_Y^X = c_X x_X x_Y, f_Y^Z = c_Z x_Y x_Z$,
where $x_i = N_i/V$$(i = X, Y, Z)$ and $N_i$ is the number of molecule species $i$.
The rates of replicating $X$ and $Z$ are, respectively, given by 
\[ F_X = V f_X s_X, \hspace{5mm} F_Z = V f_Z s_Z\]
Here, $f_X = c_Y x_X x_Y,$ and $f_Z = c_Y x_Y x_Z$.
$s_i = S_i/S_i^0$$(i=X, Y, Z)$ is the normalized concentration of the resource where $S_i$ is the concentration and $S_i^0$ is introduced to normalize $S_i$ to one when $S_i = S_i^0$.

The rate equation for molecule species $Y$ is written as, 
\begin{equation}
\frac{dN_Y}{dt} = F_Y = V (f_Y^X + f_Y^Z) s_Y , 
\label{I-1}
\end{equation}
and those for $X$ and $Z$ are written as, 
\begin{equation}
\frac{dN_X}{dt} = F_X = V f_X s_X, \frac{dN_Z}{dt} =  F_Z = V f_Z s_Z. 
\label{I-2}
\end{equation}
The dynamics of resources $S_X$, $S_Y$, and $S_Z$ are respectively written as
\begin{equation*}
\frac{dS_Y}{dt} = -  V \left( f_Y^X + f_Y^Z \right) s_Y +DS_Y^0(1 - s_Y), 
\end{equation*}
\begin{equation*}
\frac{dS_X}{dt} = - V f_X s_X  +DS_X^0(1 - s_X), 
\end{equation*}
\begin{equation*}
\frac{dS_Z}{dt} =  - V f_Z s_Z +DS_Z^0(1 - s_Z).
\end{equation*}

In the steady state, the value of $s_Y$, $\bar{s}_Y$, is written as,
\begin{equation*} 
\bar{s}_Y = \frac{DS_Y^0}{ V \left(f_Y^X + f_Y^Z \right) + DS_Y^0}.
\label{I-6}
\end{equation*}
For large $D$, $\bar{s}_Y \rightarrow 1$.
As $D$ is decreased, $\bar{s}_Y$ starts to decrease and deviates from 1. 
For smaller $D$, $\bar{s}_Y$ decreases linearly as $\alpha D$. 

Similarly, for resources $S_X$ and $S_Z$ the steady state values are written as
\begin{equation*} 
\bar{s}_X = \frac{DS_X^0}{V f_X + DS_X^0}, \hspace{5mm} \bar{s}_Z = \frac{DS_Z^0}{V f_Z + DS_Z^0}.
\label{I-7}
\end{equation*}

We assume here that the values of resource are fixed to the steady-state values.
Then, 
\[ F_Y = \frac{V (f_Y^X+f_Y^Z) D S_Y^0}{V (f_Y^X+f_Y^Z)+DS_Y^0}, \]
\[ F_X = \frac{V f_X D S_X^0}{V f_X + DS_X^0},  F_Z = \frac{V f_Z D S_Z^0}{V f_Z + DS_Z^0}.\]

Here, the dynamics of each molecule change with the diffusion constant $D$. 
When $D$ is sufficiently large, each term approaches $F_Y \rightarrow V f_Y$, and 
$F_X \rightarrow V f_X, F_Z \rightarrow V f_Z$.
On the other hand, as $D$ decreases, each term approaches 
$F_Y \rightarrow D S_Y^0$, and $F_X \rightarrow DS_X^0$, $F_Z \rightarrow DS_Z^0$.

We assumed that the volume $V$ was constant such that 
\begin{equation*}
\frac{dN_\sigma}{dt} = F_\sigma  - x_\sigma (F_Y + F_X + F_Z) , 
\label{I-8}
\end{equation*}
where $\sigma = X, Y, Z$.
In a steady state, the following equations hold,
\[ F_\sigma = x_\sigma (F_Y + F_X + F_Z).\] 
The equations are explicitly written as 
\begin{equation*}
\frac{(c_X x_X + c_Z x_Z)S_Y^0}{(c_X x_X + c_Z x_Z) N_Y + DS_Y^0} = x_Y \left( \frac{c_Y x_X S_X^0}{c_Y x_X N_Y + DS_X^0}+\frac{c_Y x_Z S_Z^0}{c_Y x_Z N_Y + DS_Z^0}+ \frac{(c_X x_X + c_Z x_Z ) S_Y^0}{N_Y (c_X x_X+ c_Z x_Z )+DS_Y^0} \right), 
\label{I-10}
\end{equation*} 
\begin{equation*}
\frac{c_Y x_X S_X^0}{c_Y x_X N_Y + DS_X^0} = x_X \left( \frac{c_Y x_X S_X^0}{c_Y x_X N_Y + DS_X^0}+\frac{c_Y x_Z S_Z^0}{c_Y x_Z N_Y + DS_Z^0}+ \frac{(c_X x_X + c_Z x_Z ) S_Y^0}{N_Y (c_X x_X+ c_Z x_Z )+DS_Y^0} \right), 
\label{I-11}
\end{equation*}  
\begin{equation*}
\frac{c_Y x_Z S_Z^0}{c_Y x_Z N_Y + DS_Z^0} = x_Z \left( \frac{c_Y x_X S_X^0}{c_Y x_X N_Y + DS_X^0}+\frac{c_Y x_Z S_Z^0}{c_Y x_Z N_Y + DS_Z^0}+ \frac{(c_X x_X + c_Z x_Z ) S_Y^0}{N_Y (c_X x_X+ c_Z x_Z )+DS_Y^0} \right). 
\label{I-12}
\end{equation*}

In Fig. 2 in the main text, we show nullclines for Eqs. (\ref{I-1}) and (\ref{I-2}) for parameters $c_X = c_Y = c_Z = 1$ and $S_X^0 = S_Y^0 = S_Z^0 = 1$, $V=1$ and $x_X + x_Y + x_Z = 1$. 
For $D = 0$, the fixed point $(x_X, x_Y, x_Z )=(1/3, 1/3, 1/3)$ is stable. As $D$ increases, the three nullclines change and finally fall into a single line. 
On the single line, the system is neutral and undergoes a random walk to reach either of the dominance points $(x_X, x_Y, x_Z)=(1/2, 1/2, 0)$ and $(0, 1/2, 1/2)$.

Explicit forms of the solutions are generally complicated. Therefore, we consider here simpler cases in which the diffusion constant $D$ is large or small.
For small $D$ values, i.e., for $(c_X x_X + c_Z x_Z) N_Y  \gg DS_Y^0$, $c_Y x_X N_Y  \gg DS_X^0$ and $c_Y x_Z N_Y  \gg DS_Z^0$, the coexistence solution exists as 
\[ x_Y = \frac{S_Y^0}{S_X^0 + S_Z^0 + S_Y^0}, x_X = \frac{S_X^0}{S_X^0 + S_Y^0 + S_Z^0}, x_Z = \frac{S_Z^0}{S_X^0+S_Y^0+S_Z^0}.\]

On the other hand, in the limit of large $D$ values, i.e., for $(c_X x_X + c_Z x_Z) N_Y  \ll DS_Y^0$, $c_Y x_X N_Y  \ll DS_X^0$ and $c_Y x_Z N_Y  \ll DS_Z^0$, 
\begin{equation*}
(c_X x_X + c_Z x_Z) = x_Y \left\{ ( c_Y + c_X ) x_X + (c_Y + c_Z) x_Z \right\}, 
\end{equation*}
\begin{equation*}
c_Y x_X = x_X \left\{ (c_Y + c_X) x_X + (c_Y + c_Z) x_Z \right\},
\end{equation*}
\begin{equation*}
c_Y x_Z = x_Z \left\{ (c_Y + c_X) x_X + (c_Y + c_Z) x_Z \right\}, 
\end{equation*}
therefore,
\[ (x_X, x_Y, x_Z) = \left( \frac{c_Y}{c_X + c_Y}, \frac{c_X}{c_X + c_Y}, 0 \right), \left(0, \frac{c_Z}{c_Z+c_Y}, \frac{c_Y}{c_Z+c_Y} \right).\]

\section{Discussion of the estimate of the optimum number of molecule species}

The number of molecule species in a cell increases as the resources are limited.
Therefore, the number of catalysts also increases for each species because catalysts are randomly assigned for each species from the possible $K_M$ species in the model.
In this section, we discuss the effects of increase of the number of catalysts with the number of molecule species 
in estimating the optimum number of species; the effect is neglected in the main text.
To estimate how increase in the number of species in a cell changes the growth speed, 
we considered the number of existing species, $K^*_M$, out of $K_M$ possible molecule species. 
Here, we assumed that $K_M$ was sufficiently large to assure that cells could increase $K^*_M$ to the optimum value.
We consider a fixed number of molecules with $K^*_M$$(\leq K_M)$ molecule species in which 
the $K^*_M$ species mutually catalyze the replication of each other.
The increase in the number of a species $X_i$, $N_i$, in the molecules is written as
$\frac{dN_i}{dt} \sim \sum_j x_i x_j S_i$. 
Here, $x_{i(j)}$ is the concentration of species $X_i(X_j)$, and the term is summed for 
all the catalysts $X_j$ for $X_i$ in the present $K^*_M$ species. 
If we assume that the concentrations of $K^*_M$ molecules are approximately the same, 
the concentration $x_{i(j)}$ decreases as $\sim 1/K^*_M$ 
with increasing $K^*_M$(up to $K_M$), if the number of all the molecules is fixed. 
Therefore the increase is estimated as 
\begin{equation*}
\frac{dN_i}{dt} \sim \sum_{j} \frac{1}{K^{*2}_M} S_i \approx \frac{C(i, K^*_M)}{K^{*2}_M} S_i,
\label{dN}
\end{equation*}
where the sum $j$ is taken for all the catalysts $X_j$ of species $X_i$.
Here $C(i, K^*_M)$ represents the number of catalysts of species $X_i$ in the $K^*_M$ species.

The dynamics of a resource $S_i$ are then written as 
\[ \frac{dS_i}{dt} \sim -\frac{C(i, K^*_M)}{K^{*2}_M} S_i + D(S_i^0 - S_i), \]
and in a steady state
\[ \bar{S_i} = \frac{DS_i^0}{C(i, K^*_M)/K^{*2}_M + DS_i^0}. \]

If we assume that $S_i^0 = S^0$ for all $i$ and $C(i, K^*_M)$ is approximately independent on each $i$, the growth rate $G$ of the cell is
\[ G = \sum_i \frac{dN_i}{dt} \sim \frac{K^*_M D S^0}{1+ DS^0 K^{*2}_M/C(K^*_M)}. \]

To determine how $C(K^*_M)$ increases, we examined the average number of catalyst for major species as a function of the number of major species (Fig.\ref{fig:S2a}). 
The three-member cycle corresponds the point $(3,1)$ in the figure. 
As $D$ decreases, the number of species increases so that the points distributed to the right.
When the number of species is between $3$ to $10$, the average number of catalysts 
is distributed rather broadly between one and two. 
This result suggests that the number of catalysts does not strongly depend on the number of species 
in the range, supporting the basic $D^{-1/2}$ scaling.

As the number of species further increases, corrections to the scaling are needed.  
While the functional form of the increase is not conclusive from the data, we note the following corrections.
If $C(K^*_M)$ increases as $C(K^*_M) \approx \log(K_M^*)$, 
the logarithmic correction as $K^{*2}_M/\log(K^*_M) \approx 1/D$ is needed.  
If we assume that the function $C(K^*_M)$ is in the form $C(K^*_M) = K_M^{*\alpha}$, a correction from the power $K_M^* \sim D^{-1/2}$ is applied.
For $\alpha < 1$, the optimum $K^*_M$ scales as $\left\{ \frac{1}{DS^0(1-\alpha)} \right\}^{1/(2-\alpha)}$ for smaller $D$ values. 
When the increase in the number of catalysts is negligible, i.e., $\alpha = 0$, the scaling $D^{-1/2}$ is reproduced. 
By approximating the increase based on the power $\alpha = 1/2$ in Fig. \ref{fig:S2a}, 
the diversity increases with $D^{-2/3}$. The slope is shown in Fig. \ref{fig:S2b}.

\section{A case of one-dimensional chain structure of mutually catalytic molecules}
We consider mutually catalytic reactions, 
\[ X_i + X_{i+1} + S_i \rightarrow 2X_i + X_{i+1}, \hspace{5mm} X_{i+1} + X_i + S_{i+1} \rightarrow 2X_{i+1} + X_i\] 
$(i = 1, ..., K_M)$ with periodic boundary, i.e., $X_{K_M+1}$ denotes $X_{1}$.
We refer to the above reactions with $X_i$ and $X_{i+1}$ as $i$-th reactions.

The rate equation for molecule species $X_i$ was written as, 
\begin{equation*}
\frac{dN_{i}}{dt} = (f_i^{i-1} + f_i^{i+1}) s_i , 
\label{c1}
\end{equation*}
where $f_i^{i-1} = c_{i-1} x_{i-1} x_i$, and $f_i^{i+1} = c_{i+1} x_{i+1} x_i$. 
Here $x_i$ denotes $N_i/N$ where $N=\sum_i N_i$.  
The dynamics of resources $S_i$ are written as
\begin{equation*}
\frac{dS_i}{dt} = -  \left( f_i^{i-1} + f_i^{i+1} \right) s_i +DS_i^0(1 - s_i), 
\end{equation*}
In a steady state, 
\[ s_i = \frac{DS_i^0}{(f_i^{i-1}+f_i^{i+1})+DS_i^0}. \]

Here, we assume $c_i = 1$ for all $i$. 
For the minimum case when a pair of $X_i$ and $X_{i+1}$ exists in the system, $N_i = N_{i+1} = N/2$(i.e.$x_i = x_{i+1} = 1/2$), and $f_i^{i-1} = 0$, $f_i^{i+1} = 1/4$ thus, $s_i = \frac{DS_i^0}{1/4 + DS_i^0}.$
Then, the increase is written as 
\begin{equation*}
\frac{dN}{dt} = \frac{dN_i}{dt} + \frac{dN_{i+1}}{dt} = \frac{DS_i^0}{1+4DS_i^0} +  
\frac{DS_{i+1}^0}{1 +4DS_{i+1}^0}.
\end{equation*}

For cases in which more molecule species coexists in the system, we developed the following explanation.
For a general case where $n$ species coexist($2 \leq n \leq K_M$), $N_{i-n/2+1} = ... = N_{i} = N_{i+1} = ... = N_{i+n/2} = N/n$(i.e. $x_j = 1/n$), and
$f_j^{j-1} = 1/n^2$, $f_j^{j+1} = 1/n^2$ thus, $s_j = \frac{DS_j^0}{2/n^2 + DS_j^0}$ except that $j = i-n/2+1, i+n/2$.
For $i-n/2+1$ and $i+n/2$, $s_j = \frac{DS_j^0}{1/n^2 + DS_j^0}$.
Thus, 
\begin{equation*}
\frac{dN}{dt} = \sum_j \frac{dN_j}{dt} = \frac{1}{n^2} \frac{DS_{i-n/2+1}^0}{1/n^2 + DS_{i-n/2+1}^0} + 
\sum_{j=i-n/2+2}^{i+n/2-1} \frac{2}{n^2} \frac{DS_j^0}{2/n^2 + DS_j^0} + \frac{1}{n^2} \frac{DS_{i+n/2}^0}{1/n^2 + DS_{i+n/2}^0}.
\end{equation*}

For a case $S_{j}^0 = S^0$ for all $j$, the increase $dN/dt$ is written as
\begin{equation*}
\frac{dN}{dt} = \frac{2DS^0}{1+n^2DS^0} + \frac{(n-2)DS^0}{1+n^2DS^0/2}.
\end{equation*}
Here, we note that the first term denotes the increase of the two species located at both ends of the successive indices, and 
the second term denotes the sum of the intermediate species (for $n=2$, there is only the first term). 
For a more general case in which $n$ species coexist in the system, but the species can be divided into subgroups, e.g., 
first group($X_i$, $X_{i+1}$, ..., $X_{i+n_1}$) and second group($X_j$, $X_{j+1}$,..., $X_{j+n_2}$) and so on($n=\sum n_i$).
The estimate will be modified such that more species will be categorized into the first term and less species will be categorized into the second term.
Such subgroups were actually observed in our stochastic simulation(Fig. \ref{fig:3}), 
because, with $\mu = 0$, some species that were lost by stochasticity in division events never appeared again in the system. 
Our estimate suggests, however, that $n$ species in a single group(successive indices) will give the maximum of $dN/dt$ because $DS^0/(1+n^2DS^0) < DS^0/(1+n^2DS^0/2)$.

To estimate the optimum number of species $n^*$ to achieve the maximum growth speed, we consider a function $f(x; n) = 2x/(1+n^2x) + (n-2)x/(1+n^2x/2)$.
Here, the increase $dN/dt$ is written as $f(DS^0; n)$. Then the optimum $n^*$ satisfies the 
condition, given a fixed $x=DS^0$, 
\begin{equation*}
\frac{df(x; n)}{dn} = -\frac{(n-2)nx^2}{(1+n^2x/2)^2} + \frac{x}{1+n^2x/2} - \frac{4nx^2}{(1+n^2x)^2} = 0.
\end{equation*}
The red curve in Fig. \ref{fig:3} shows a numerical solution for $n^*$ to satisfy the condition. 
For a small $x$, 
\[ \frac{df(x; n)}{dn} = -(n-2)nx^2 + x - 4nx^2 = -(n+2)nx^2 + x \]
$df/dn=0$ reads $(n^*+2)n^* = 1/x$. This leads to the dependence $n^* \approx (DS^0)^{-1/2}$ for $DS^0 \rightarrow 0$.

\newpage
\section{Supplementary figures}


\begin{figure}[htbp]
  \begin{center}
    \begin{tabular}{c}
      \begin{minipage}{0.5\hsize}
        \begin{center}
        \rotatebox{270}{
          \includegraphics[clip, width=6cm]{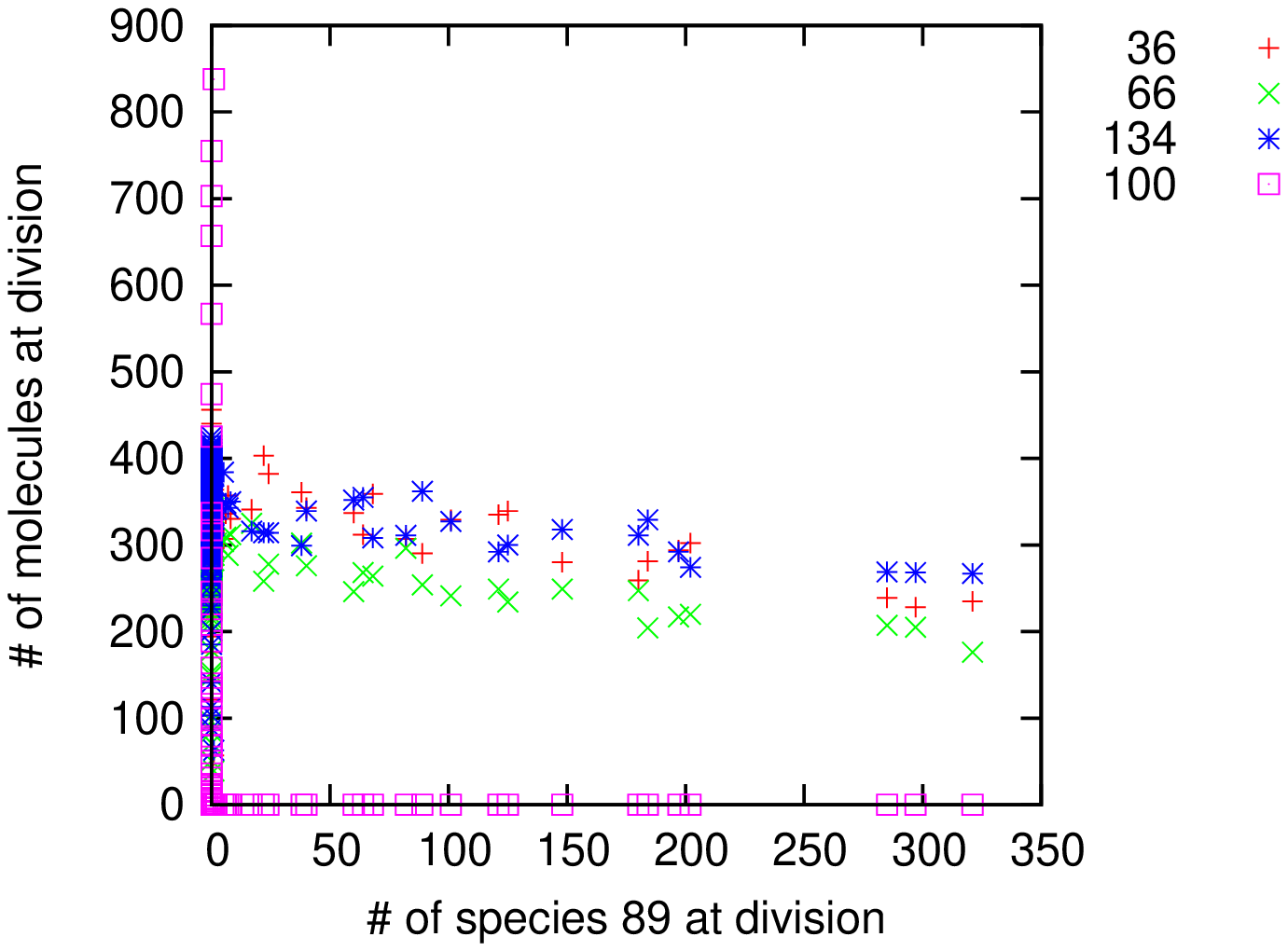}}
        \end{center}
      \end{minipage}
      
      \begin{minipage}{0.5\hsize}
        \begin{center}
        \rotatebox{270}{
          \includegraphics[clip, width=6cm]{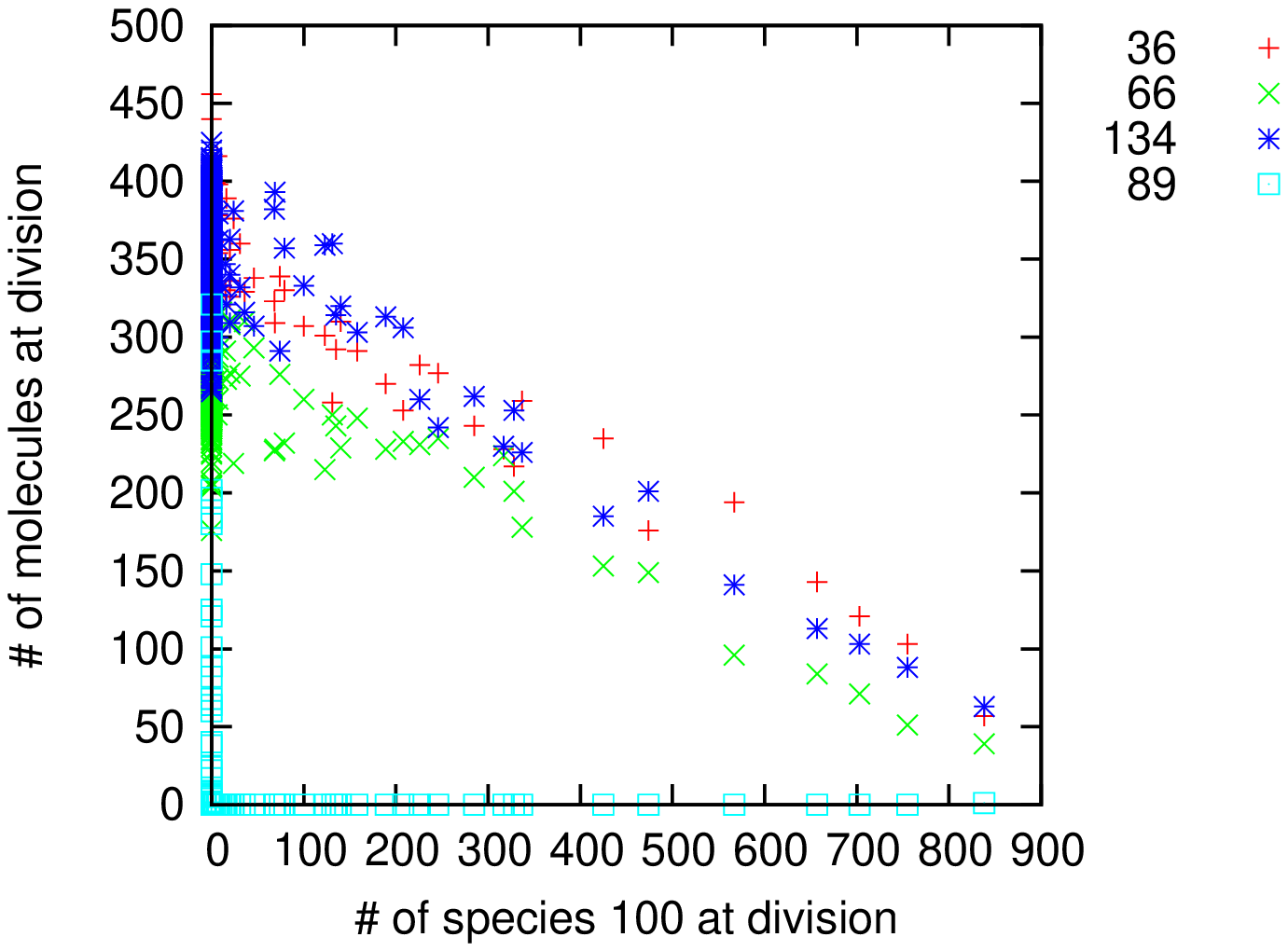}}
        \end{center}
      \end{minipage}
      
     \end{tabular}
     \caption{The number of molecule species as a function of those of species 89(left) and species 100(right) at the division events shown in Fig. 1(A)(a) of the main text.
     Cells with more parasite species 89 or 100 contain fewer hypercycle members(36, 66, and 134).}
 \end{center}
\end{figure}

\begin{figure}[htbp]
  \begin{center}
    \begin{tabular}{c}
      \begin{minipage}{0.5\hsize}
        \begin{center}
        \rotatebox{270}{
          \includegraphics[clip, width=6cm]{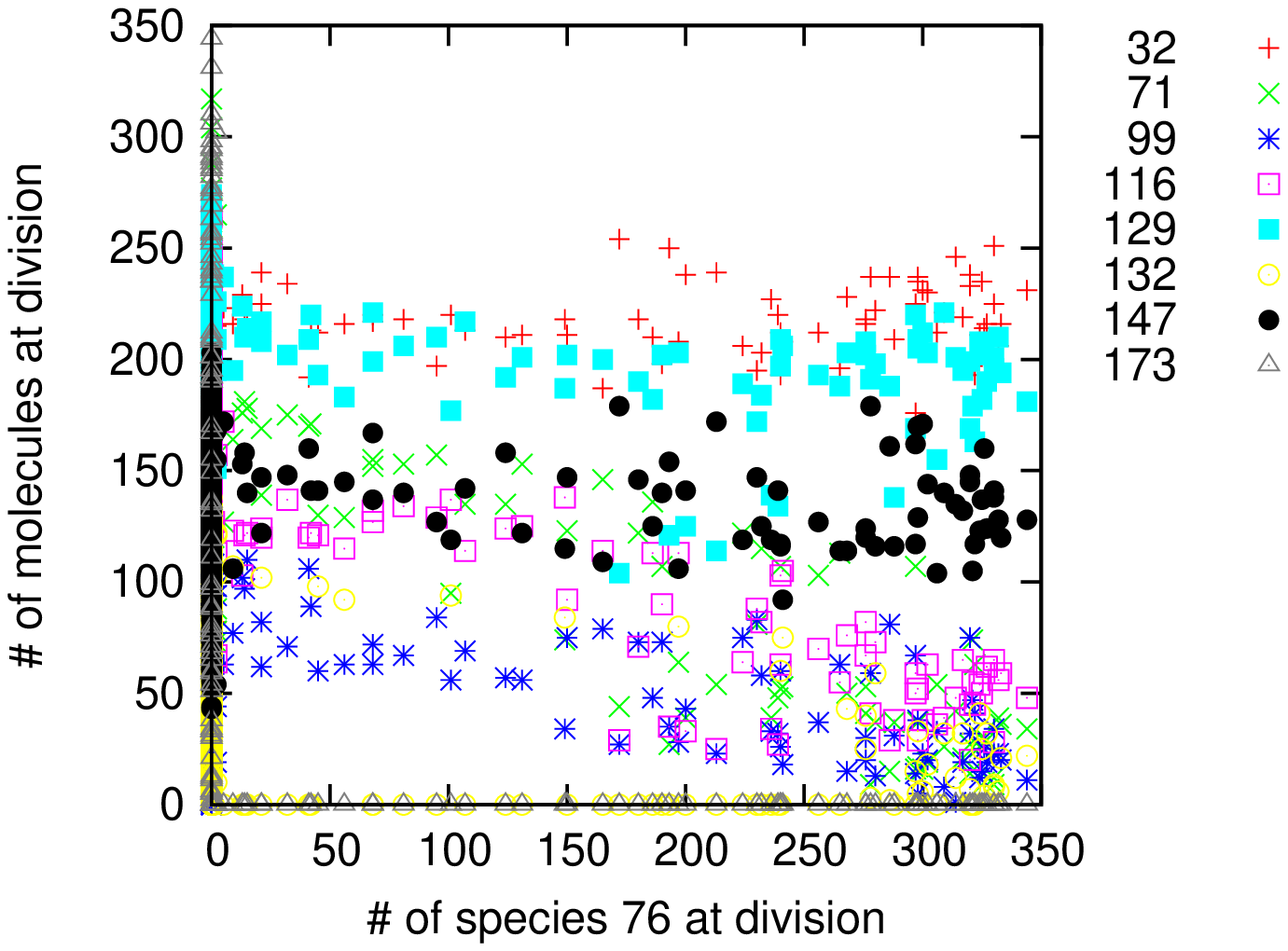}}
        \end{center}
      \end{minipage}
      
      \begin{minipage}{0.5\hsize}
        \begin{center}
        \rotatebox{270}{
          \includegraphics[clip, width=6cm]{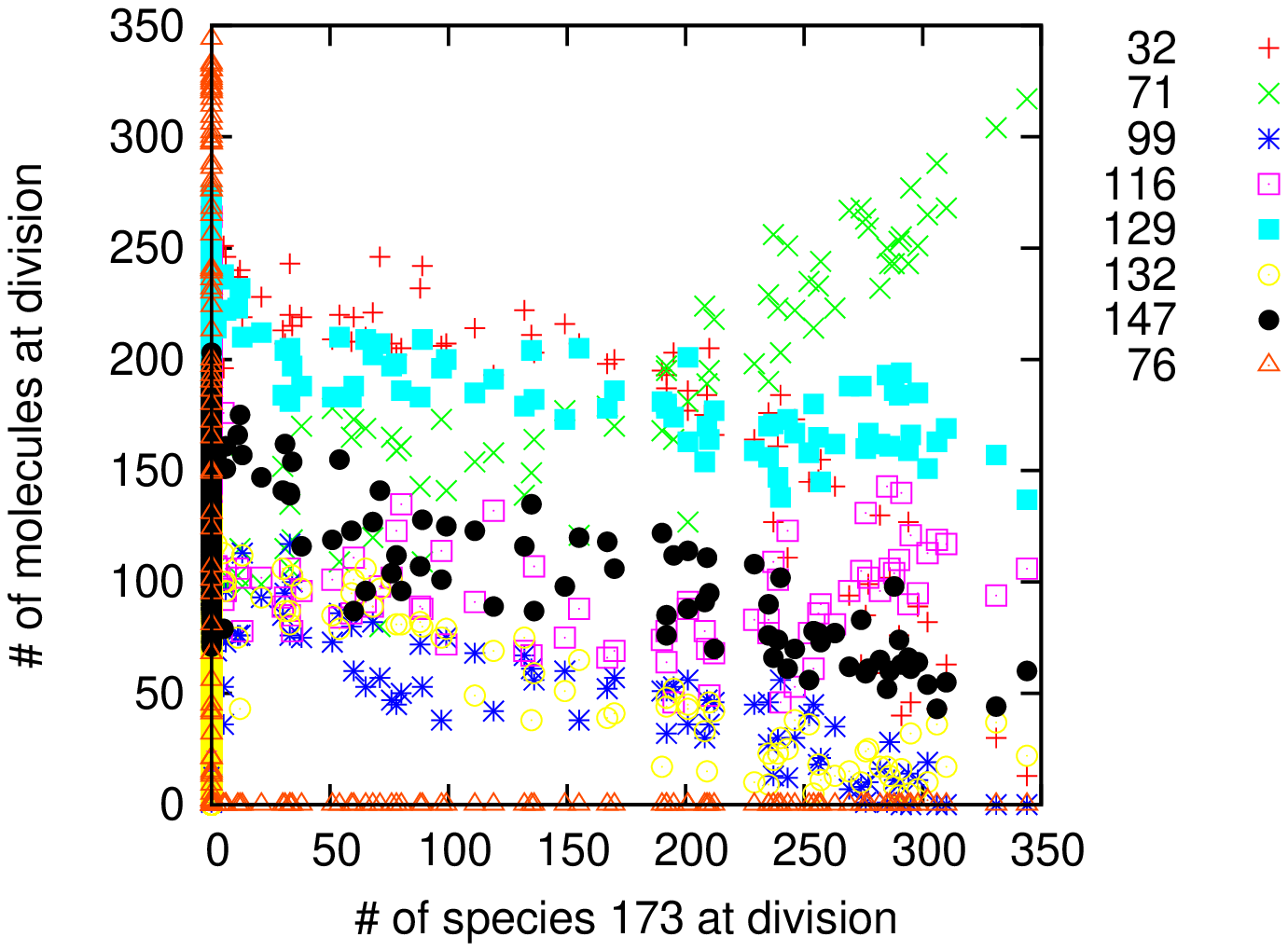}}
        \end{center}
      \end{minipage}
      
     \end{tabular}
     \caption{The number of molecule species as a function of those of species 76(left) and species 173(right) at the division events shown in Fig. 1(A)(b) of the main text.
     For cells with parasite species 76, species 32, 129 and 147 are relatively preserved to produce the parasite, while the numbers of the others decrease remarkably.
     For cells with increased numbers of species 173, all the other species decrease, except species 71, which replicates with the aid of species 173.}
 \end{center}
\end{figure}

\newpage 

\begin{figure}[t]
\begin{center}
\rotatebox{270}{
\includegraphics[width=8cm, clip]{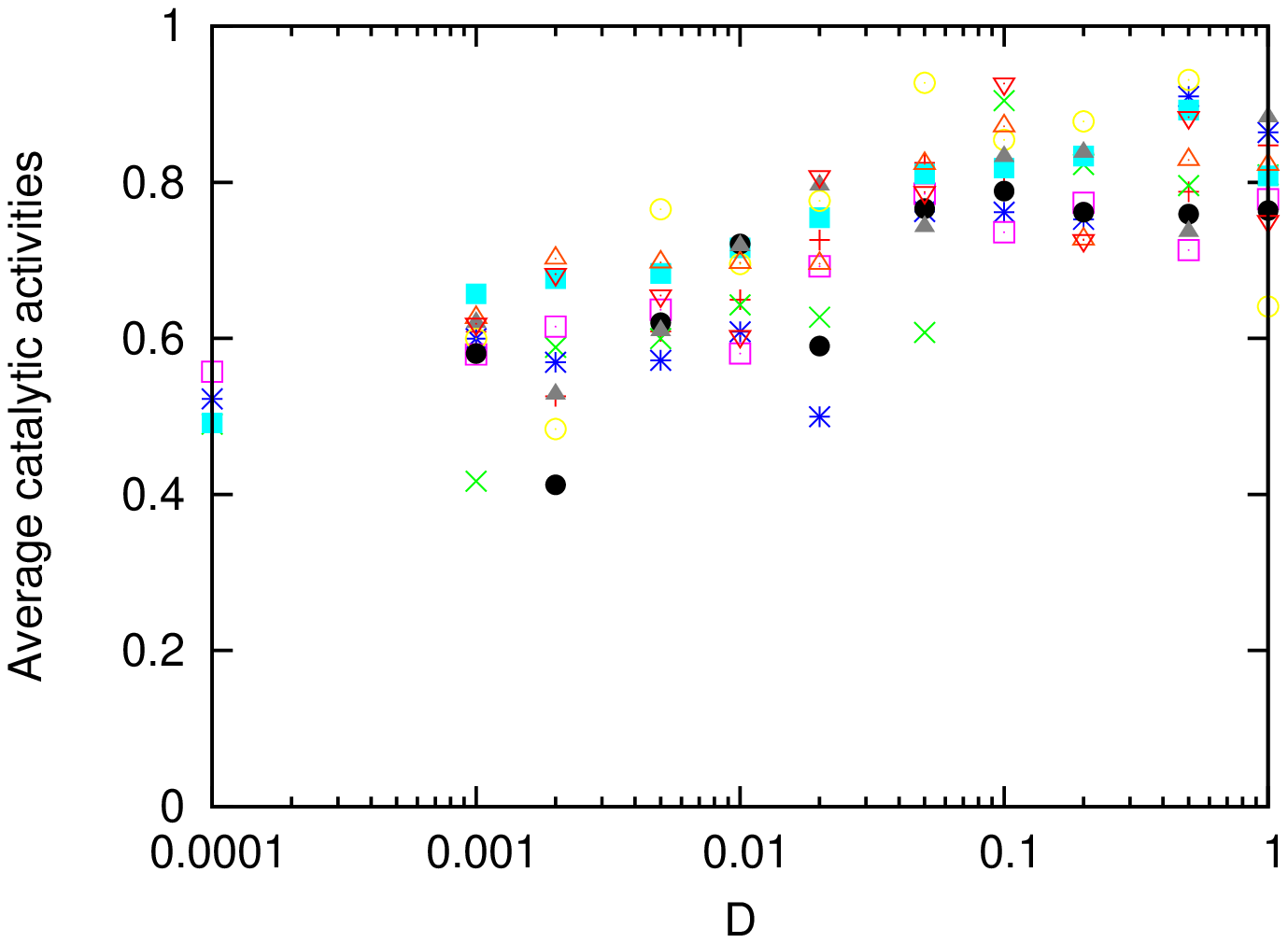}}
\vskip 0.25cm
\caption{The average catalytic activity $c_i$. Each point corresponds to the average value of $c_i$ of the major molecule species. 
Different colors indicate different network samples. Parameters are $N=1000$ and $\mu = 0.001$.}
\label{fig:S4}
\end{center}
\end{figure}

\begin{figure}[t]
\begin{center}
\rotatebox{270}{
\includegraphics[width=8cm, clip]{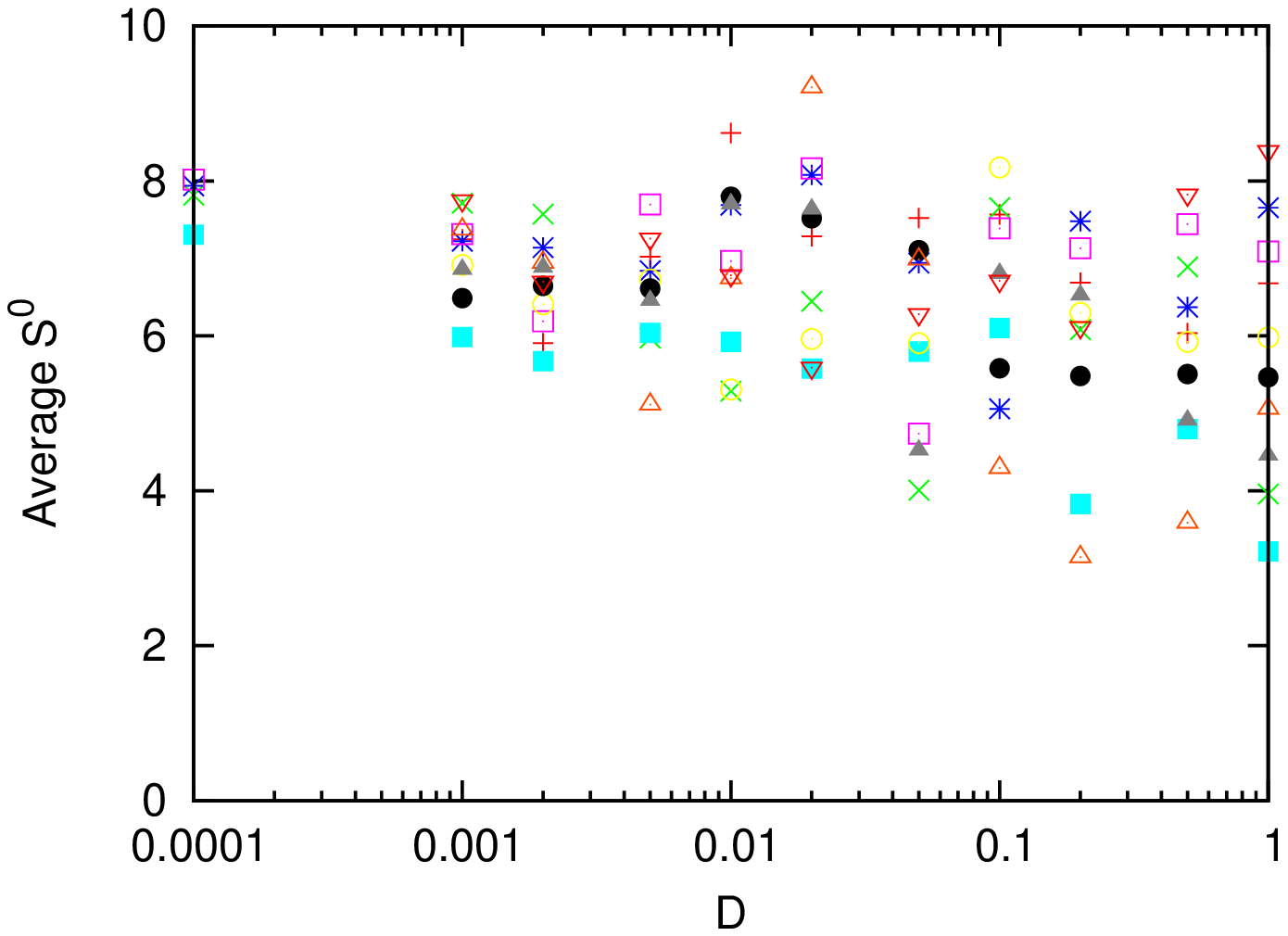}}
\vskip 0.25cm
\caption{The average values of resource reservoir $S^0_i$. Each point corresponds to the average value of $S_i^0$ of the major molecule species. 
Different colors indicate different network samples. Parameters are $N=1000$ and $\mu = 0.001$.}
\label{fig:S5}
\end{center}
\end{figure}

\begin{figure}[t]
\begin{center}
\rotatebox{270}{
\includegraphics[width=6cm]{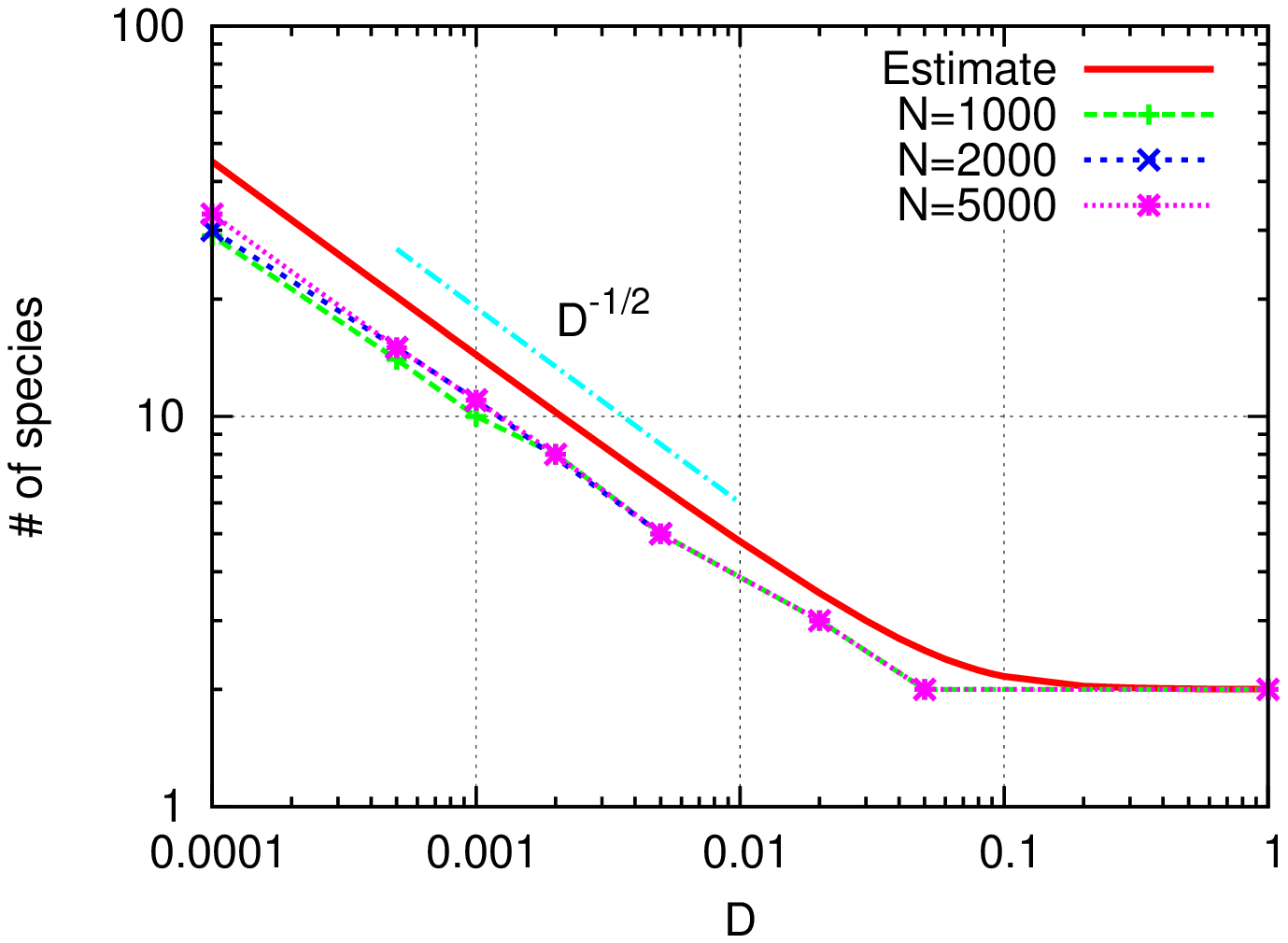}}
\vskip 0.25cm
\caption{The number of species as a function of $D$ in the linear-chain network model for $N= 1000,2000$, and $5000$. 
Here, $c_i = 1$, $S_i^0 = 10$ for all $i$$(i=1,...,K_M)$, and $\mu = 0$. The red curve shows numerical estimation based on solving rate equations(see $\S$4). 
The slope $D^{-1/2}$ is also shown.
In this case, the minimum is a two-species hypercycle and the number transit to increase around the estimated balance point $D^* = c_i/4S_j^0 = 0.025$.
}
\label{fig:3}
\end{center}
\end{figure}

\begin{figure}[htbp]
\begin{center}
\begin{tabular}{c}
\begin{minipage}{0.5\hsize}
\rotatebox{270}{
\includegraphics[width=5cm]{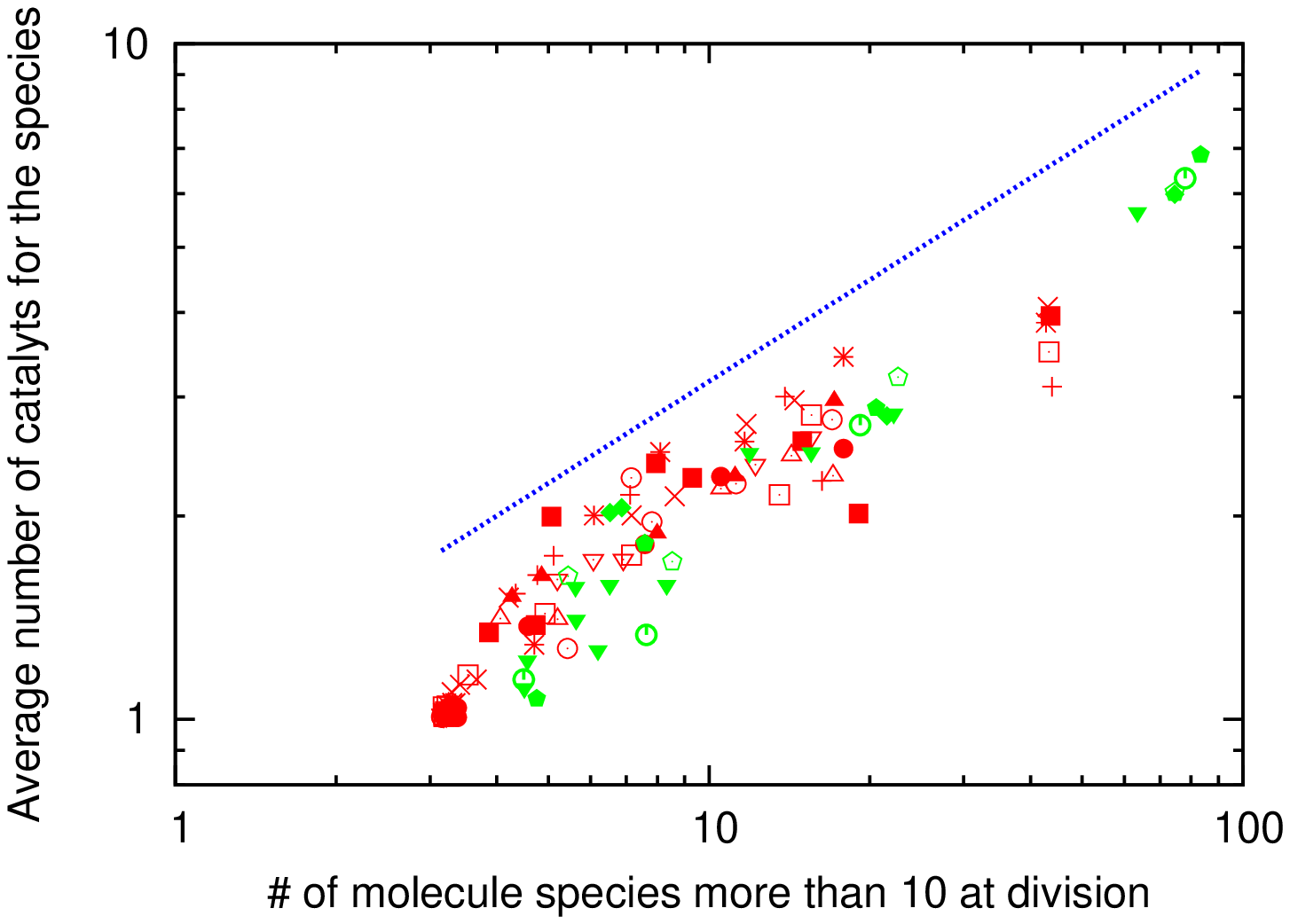}}
\hspace{1.5cm} (a)
\end{minipage}
\begin{minipage}{0.5\hsize}
\rotatebox{270}{
\includegraphics[width=5cm]{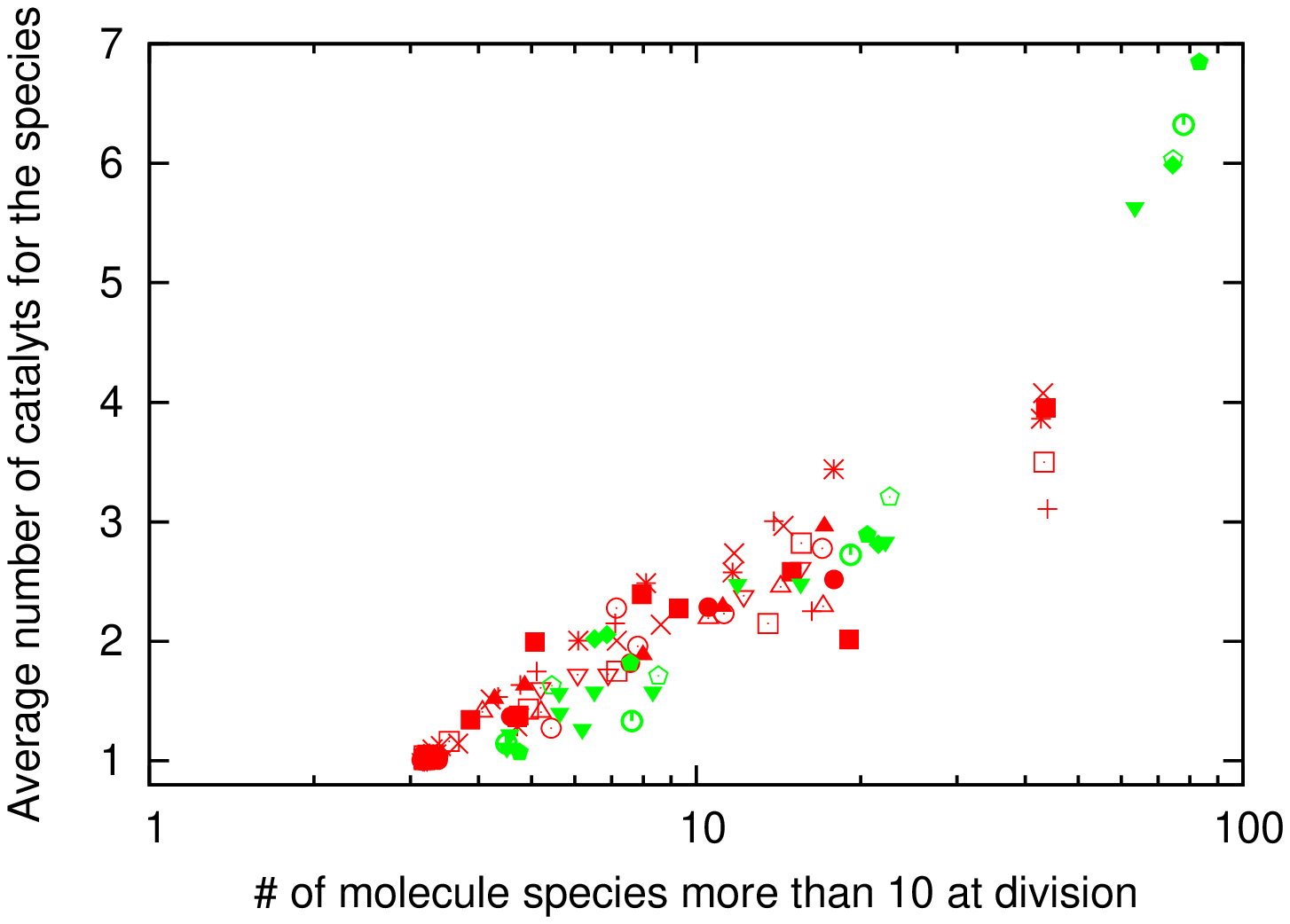}}
\hspace{1.5cm} (b)
\end{minipage}
\end{tabular}
\end{center}
\caption{The average number of existing catalyst species for major species plotted as a function of 
the number of major species in (a)log-log and (b)semi-log scales.
The red, and green points show the results for 
N = 1000, and 5000, respectively, while different symbols show results of different network samples with the network path probability fixed to $\rho = 0.2$.
For each example(symbols), data for $D = 1, 0.5, 0.2, 0.1, 0.05, 0.02, 0.01, 0.005, 0.002, 0.001$ and $0.0001$ are plotted.
The slope $x^{1/2}$ is also shown for (a).
}
\label{fig:S2a}
\end{figure}

\begin{figure}[htbp]
\begin{center}
\rotatebox{270}{
\includegraphics[width=6cm]{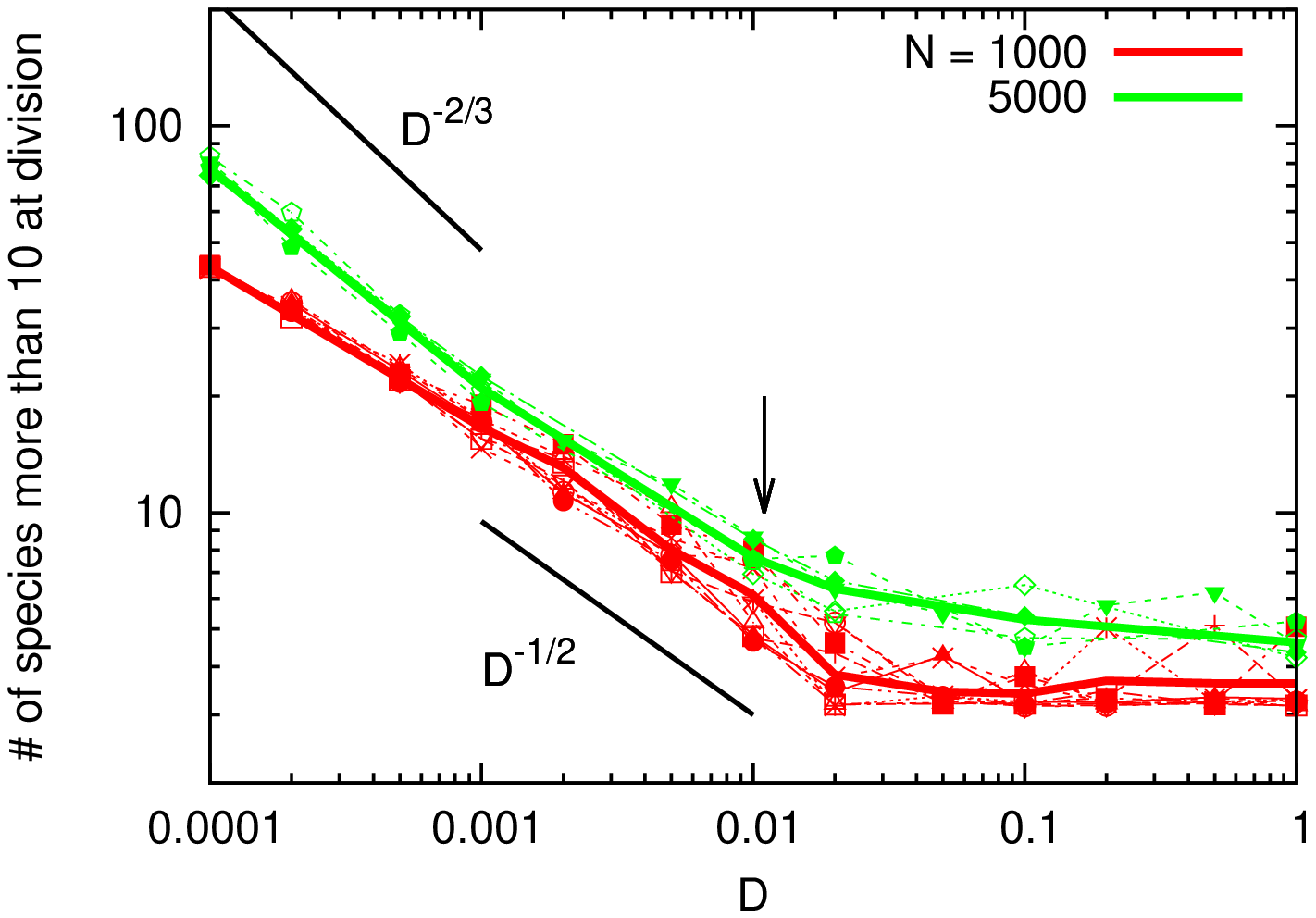}}
\vskip 0.25cm
\caption{The number of major species(more than 10 copies at division) as a function of D for N = 1000 and 5000.}
\label{fig:S2b}
\end{center}
\end{figure}